\documentstyle[fleqn,12pt]{article}

\def\thalf{{\textstyle{\frac{1}{2}}}}
\def\threehalf{{\textstyle{\frac{3}{2}}}}
\def\tquar{{\textstyle{\frac{1}{4}}}}
\def\threequar{{\textstyle{\frac{3}{4}}}}
\def\oneth{{\textstyle{\frac{1}{3}}}}
\def\onetwelfth{{\textstyle{\frac{1}{12}}}}

\def\fourth{{\textstyle{\frac{4}{3}}}}
\def\lord{$ \raisebox{-.3ex}{$\stackrel{<}{_{\sim}}$} $}
\def\pj{\hspace{-.26cm}}

\begin{document}
\title{Numerical Approximation to the Thermodynamic Integrals}
\author{S.M. Johns\thanks{Present address: Department of Physics,
 Cornell University, Ithaca NY  14853.}\,\,  and P.J. Ellis \protect\\
{\small School of Physics and Astronomy, University of Minnesota,}
\protect\\{\small Minneapolis, MN 55455-0112}
\protect\\J.M. Lattimer \protect\\{\small Department of Earth and Space
Sciences, SUNY at Stony Brook,}
\protect\\{\small Stony Brook, NY 11794-2100}}
\date{~}
\maketitle
\thispagestyle{empty}
\pagestyle{myheadings}
\markboth{~}{~}
{\small
\centerline{{\bf Abstract}}
We approximate boson thermodynamic integrals as polynomials in two variables
chosen to
give the correct limiting expansions and to smoothly
interpolate into other regimes. With 10 free parameters, an accuracy of
better than 0.009\% is achieved for the pressure, internal energy density and 
the number density.   We also revisit the fermion case,
originally addressed by Eggleton, Faulkner and Flannery (1973), and
substantially improve the accuracy of their fits.}
\vskip-20.5cm
\hfill NUC-MINN-95/22-T\\
\newpage
\noindent{\small{\it Subject headings:} thermodynamics -- numerical analysis --
bosons, fermions}
\section {INTRODUCTION}

The thermodynamic functions of fermions and bosons are expressed as
integrals which cannot be evaluated analytically except in limiting
cases of degeneracy or relativity. Numerical integration is not entirely
straightforward, however, since the form of the integrand varies widely 
under different conditions. Thus the numerical method chosen
must be carefully tested and optimized for the parameters at hand and 
different methods may be needed in different regimes.
Furthermore, the computational time
required can be an issue if the integrals need to be evaluated many
times. For fermions, these problems
can be  overcome by using an ingenious numerical approach described by
Eggleton, Faulkner, \& Flannery (1973), hereafter referred to as EFF.
They approximate the integrals by means of a polynomial in two
variables. The polynomial form is chosen so that it yields the correct
behavior in the four limiting situations for which  the series can be
derived analytically. The coefficients are then optimized to accurately 
interpolate into regimes which are inaccessible to analytical analysis.
We have found that a simple  modification of the approach of EFF,
which involves a  single additional parameter, can
significantly improve the accuracy for a given number of coefficients
and, in addition, permits the approximation to merge exactly into the
correct asymptotic behavior. This modification is superior to the one
discussed by Pols {\it et al.} (1995).  We discuss the fermion case
briefly in \S 3.

The main purpose of the present paper is to analyze the thermodynamic
integrals for bosons; any
condensate contribution would, of course, have to be handled separately.
 While we follow the general  approach of EFF, bosons
require a different approximation  scheme.  This is discussed in \S 4,
in which we also present results for various orders of
approximation.  

We begin by defining the integrals at hand in \S 2.  The fermion and
boson cases are discussed in \S 3 and 4, respectively, and brief
concluding remarks are given in \S 5.

\section{DEFINITIONS}

For a relativistic system of particles, the standard form for the pressure is
\begin{equation}
P=\pm gT\int\frac{d^3k}{(2\pi\hbar)^3}\ln\left[1\pm e^{-(E_k-\mu)/T}\right]\;,
\label{pdef}
\end{equation}
where $g$ is a degeneracy factor, the energy $E_k=\sqrt{k^2c^2+m^2c^4}$, $\mu$
is the chemical potential and the upper (lower) sign refers to fermions 
(bosons). The units employed are such that Boltzmann's
constant $k_B=1$, so that the temperature is measured in units 
of energy. The angular integration is trivial for an infinite system and
it is convenient to perform the standard integration by parts. 
Expressing the result in terms of the dimensionless variables
of EFF
\begin{equation}
t=\frac{T}{mc^2}\quad;\quad\psi=\frac{\mu-mc^2}{T}\;,\label{psit}
\end{equation}
and a dimensionless integration variable $l=k/mc$, we write the pressure
in dimensionless form
\begin{equation}
p=\frac{P}{d}=\frac{1}{3}\int\limits_0^{\infty}dl\frac{l^4}
{\sqrt{l^2+1}}\left[e^{\frac{E}{t}-\psi}\pm1\right]^{-1}\;,\label{p}
\end{equation}
where $E=\sqrt{l^2+1}-1$ and 
$d=\frac{g}{2\pi^2}mc^2\left(\frac{mc}{\hbar}\right)^3$ carries the 
dimensions of $P$.
For fermions the value of $\psi$, or equivalently $\mu$, is unrestricted.
However for bosons, since the occupation probability has to be a positive 
quantity, $\psi t=\frac{\mu}{mc^2}-1\leq0$; thus $\psi$ is always negative. 
For a gas containing antibosons ($\bar b$) in thermal equilibrium with
bosons ($b$), so that $\mu_{\bar b}=-\mu_b$, the requirement 
$\psi_{\bar{b}}t\leq0$ yields
the further restriction $\psi_b t\equiv\psi t\geq-2$.

The remaining thermodynamic functions are easily obtained. We write the 
entropy density ${\cal S}$ in dimensionless form as
\begin{equation}
s=\frac{mc^2{\cal S}}{d}=\left(\frac{\partial p}{\partial t}\right)_{\!\mu}\;.
\label{s}\end{equation}
The number density ${\cal N}$ written in dimensionless form is
\begin{equation}
\rho=\frac{mc^2{\cal N}}{d}=\frac{1}{t}\left(\frac{\partial p}{\partial\psi}
\right)_{\!t}=\int\limits_0^{\infty}dl\:l^2
\left[e^{\frac{E}{t}-\psi}\pm1\right]^{-1}\;.\label{n}
\end{equation}
The energy density ${\cal E}$  written in dimensionless form is 
\begin{equation}
e=\frac{{\cal E}}{d}=ts-p+\rho(\psi t+1)=\int\limits_0^{\infty}dl\:l^2
\sqrt{l^2+1}\left[e^{\frac{E}{t}-\psi}\pm1\right]^{-1}\;.\label{e}
\end{equation}
For $e$, the leading power series behavior of $\rho$ obviously differs from
that of $ts$, $p$ and $\rho\psi t$. We therefore work with the internal energy 
density 
\begin{equation}
u=e-\rho=t\left(\frac{\partial p}{\partial t}\right)_{\!\psi}-p\;.\label{u}
\end{equation}

\section{APPROXIMATION TO FERMION \protect\\ INTEGRALS}
\subsection{Formalism}

The approach we use follows that of EFF.
We determine an approximation for the pressure in terms of the
degeneracy parameter
$\psi$ and the temperature parameter $t$.  Then, by eqs. (\ref{s}--\ref{e}),
the remaining thermodynamic variables can be found by differentiation.
The EFF scheme requires knowledge of the expansions of
the pressure in the limiting cases.  In the non-degenerate case (ND), 
$\psi\ll-1$, the denominator in equation (\ref{p}) can be expanded in powers 
of $\exp(\psi+1/t)$ (Chandrasekhar 1958), leading to
\begin{eqnarray}
p&\pj=&\pj t^2\sum\limits_{n=1}^{\infty}(\mp 1)^{n+1}
\frac{e^{n(\psi+\frac{1}{t})}}{n^2}K_2\!\left(\frac{n}{t}\right)\;,\label{ndf}
\end{eqnarray}
where, as before, the upper sign
refers to fermions and the lower to bosons.
Here $K_2$ is a modified Bessel function which may be expanded 
(Abramowitz \& Stegun 1965) in the extremely relativistic (ER, $t\gg1$)
and non-relativistic (NR, $t\ll1$) limits.   In the extremely degenerate case 
(ED) for
fermions, in which $\psi\gg1$, the Sommerfeld expansion (Chandrasekhar 1958)
may be employed and the extremely relativistic (ER, $\psi t\gg1$),
and non-relativistic (NR, $\psi t\ll1$) limits obtained. Then the four limiting 
cases for the fermion pressure are
\begin{eqnarray}
p&\pj=&\pj \sqrt{\frac{\pi}{2}}t^{\frac{5}{2}}e^{\psi}
\sum\limits_{n=1}^{\infty}(-1)^{n+1}\frac{e^{(n-1)\psi}}{n^{\frac{5}{2}}}
\left\{1+\frac{15t}{8n}+\frac{105t^2}{128n^2}-\cdots\right\},
\,{\rm ND\ NR}\nonumber\\
p&\pj=&\pj2t^4e^{\psi}\sum\limits_{n=1}^{\infty}(-1)^{n+1}
\frac{e^{(n-1)\psi}}{n^4}\left\{1+\frac{n}{t}+\frac{n^2}{4t^2}-\cdots
\right\},\hspace{1.4cm}{\rm ND\ ER}\nonumber\\
p&\pj=&\pj\frac{4\sqrt{2}}{15}(\psi t)^{\frac{5}{2}}\left\{1+\frac{45}{64}
\psi t+\frac{15}{128}(\psi t)^2+\frac{5\pi^2}{8\psi^2}+\cdots\right\},
\hspace{1.025cm}{\rm ED\ NR}\nonumber\\
p&\pj=&\pj\onetwelfth(\psi t)^4\left\{1+\frac{4}{\psi t}+\frac{2\pi^2}
{\psi^2}+\frac{3}{(\psi t)^2}+\frac{4\pi^2}{\psi^3t}+\frac{7\pi^4}{15\psi^4}
+\cdots\right\}\mbox{,}{\rm ED\ ER}\label{flimits}
\end{eqnarray}
with the corresponding limiting cases
for the number density and internal energies easily obtained
from equation (\ref{flimits}) by differentiation.

The key to the EFF scheme is to find functions $f(\psi)$ and
$g(\psi,t)$ such that equation (\ref{flimits}) can be expressed as
\begin{eqnarray}
p&\pj= fg^{5/2}\sum\limits_{m=0}^\infty\sum\limits_{n=0}^\infty a_{mn}f^{m}g^{n}
\qquad &{\rm ND\ NR}\qquad g\ll1,\quad f\ll1\mbox{,}\nonumber\\
p&\pj= fg^4\sum\limits_{m=0}^\infty\sum\limits_{n=0}^\infty b_{mn}f^{m}g^{-n}
\qquad &{\rm ND\ ER}\qquad g\gg1,\quad f\ll1\mbox{,}\nonumber\\
p&\pj= g^{5/2}\sum\limits_{m=0}^\infty\sum\limits_{n=0}^\infty c_{mn}f^{-m}g^{n}
\qquad &{\rm ED\ NR}\qquad g\ll1,\quad f\gg1\mbox{,}\nonumber\\
p&\pj= g^4\sum\limits_{m=0}^\infty\sum\limits_{n=0}^\infty d_{mn}f^{-m}g^{-n}
\qquad &{\rm ED\ ER}\qquad g\gg1,\quad f\gg1\mbox{,}
\label{flim1}
\end{eqnarray}
where $a_{mn}, b_{mn}, c_{mn}$ and $d_{mn}$ are coefficients. 
This is possible provided that
\begin{eqnarray}
&\hspace{-2mm}\pj f(\psi)= e^\psi\sum\limits_{m=0}^\infty u_m e^{m\psi}\ \ 
\ ;\ 
g(\psi,t)= t\sum\limits_{m=0}^\infty v_m e^{m \psi}\ \ \enskip & 
{\rm ND}\quad\psi\ll-1\mbox{,}\nonumber\\
&\pj f(\psi)= \psi^2\sum\limits_{m=0}^\infty w_m\psi^{-2m}\ ;\ 
g(\psi,t)= \psi t\sum\limits_{m=0}^\infty x_m\psi^{-2m}\enskip & {\rm ED}
\quad \psi\gg1\mbox{,}
\label{flim2}\end{eqnarray}
where $u_m,\ v_m,\ w_m$ and $x_m$ are additional coefficients. Truncating 
the summations at order $M$ and $N$, equations (\ref{flim1}) can be
combined in the single expression
\begin{equation}
p=\frac{fg^{\frac{5}{2}}(1+g)^{\frac{3}{2}}}{(1+f)^{M+1}(1+g)^N}
\sum\limits_{m=0}^M\sum\limits_{n=0}^Np_{mn}f^mg^n\;,\label{fppoly}
\end{equation}
which will be used to interpolate into all regions of $\psi$ and $t$.

Equation (\ref{flim2}) indicates that $f$ should obey
\begin{eqnarray} 
f\propto e^{\psi}\ &;&\ \frac{df}{d\psi}\propto f,\qquad\psi\ll-1,
\ f\ll1\quad{\rm ND}\mbox{,}\nonumber\\
f\propto \psi^2\ &;&\ \frac{df}{d\psi}\propto\sqrt{f},
\qquad\psi\gg1,\ f\gg1\quad{\rm ED}\mbox{.}
\end{eqnarray} 
We choose for the derivative 
\begin{equation}
\frac{df}{d\psi}=\frac{f}{\sqrt{1+\frac{f}{a}}}\;,
\end{equation}
which differs from EFF by leaving
$a$ as a  free parameter to be fitted, rather than taking it to be unity
as they did.  This simple change gives a
significant improvement in the fit, because it allows greater freedom in
matching the limiting expression for the entropy in the extremely degenerate 
case. In the EFF scheme the
relative errors in the entropy in the degenerate limit remain
substantial even when the order of the approximation is very large.
Integration yields
\begin{equation}
\psi=2\sqrt{1+\frac{f}{a}}+\ln\frac{\sqrt{1+\frac{f}{a}}-1}
{\sqrt{1+\frac{f}{a}}+1}\;.
\label{psif}
\end{equation}
For the non-degenerate case, $f\simeq4ae^{\psi-2}$, which
is small; while for extreme degeneracy, $f\simeq \tquar a\psi^2$, which is 
large. We define $g=t\sqrt{1+f}$, as in EFF,
since we have found no advantage in introducing additional parameters here. 

The leading terms of equation (\ref{flimits}) yield values 
for the four ``corner" coefficients and these are collected in Table 1. In 
addition, the leading contribution to the entropy density for
the extremely degenerate case supplies constraints on the 
$p_{M-1,0}$ and $p_{M-1,N}$ coefficients. These are also listed in Table 1.

We choose the internal energy density and 
number density to be thermodynamically consistent with equation (\ref{fppoly}), 
so by differentiation we find
\begin{eqnarray}
u&\pj=&\pj\frac{fg^{\frac{5}{2}}(1+g)^{\frac{3}{2}}}{(1+f)^{M+1}(1+g)^N}
\sum\limits_{m=0}^M\sum\limits_{n=0}^Np_{mn}f^mg^n
\left[\threehalf+n+(\threehalf-N)\frac{g}{1+g}\right]\;,\nonumber\\
\rho&\pj=&\pj\frac{f[g(1+g)]^{\frac{3}{2}}}{(1+f)^{M+\frac{1}{2}}
(1+g)^N\sqrt{1+\frac{f}{a}}}\sum\limits_{m=0}^M\sum\limits_{n=0}^Np_{mn}f^mg^n
\Biggl\{1+m\nonumber\\
&&\quad+(\tquar+\thalf n-M)\frac{f}{1+f}+(\threequar-\thalf N)
\frac{fg}{(1+f)(1+g)}\Biggr\}\;.\label{funpoly}
\end{eqnarray}

\subsection{Results}

We determined the polynomial coefficients $p_{mn}$ by a least squares fit to 
data for the pressure, which was  obtained by accurate numerical integration 
using the same ($f, g$) grid as in EFF. This was done in order to facilitate 
comparisons with
EFF.  The accuracies of the number density are of the same
order as those for the pressure, so this procedure yields acceptable
results for the internal energy and number density as well. Table 2
displays the maximum modulus (MM) and root mean square (RMS) deviations
for various values of $M$ and $N$. Specifically these are defined in
terms of the fractional deviation of, for example, the pressure at the
fitting points $i$, $\Delta p_i$, according to
\begin{equation}
\Delta p_{i}=\frac{p_i^{\rm poly}-p_i^{\rm exact}}{p_i^{\rm exact}}\ ;
\ {\rm MM}={\rm Max}|\Delta p_{i}|\ ;\ 
{\rm RMS}=\sqrt{\frac{\sum_i(\Delta p_{i})^2}{\sum_i}}\;.
\end{equation}
Table 2 also shows that treating $a$ as a free parameter significantly
improves the fit as long as both $M$ and $N$ are greater than 1. For
example, for the $M=N=3 (2)$ case the improvement amounts to a factor of 
8 (4). For
fermions the number of coefficients $p_{mn}$ to be fitted is
$(N+1)(M+1)$. The $N=M=2$
case, whose coefficients are listed in Table 3, shows RMS deviations of
0.03\%, which may be sufficient for many purposes. However, an order of
magnitude improvement in the accuracy can be obtained by going to the
$N=M=3$ approximation, which is displayed in Table 4.

In many situations, it is preferable to have approximations that smoothly
merge onto the exact limits.  This prevents discontinuities from
occurring when the exact results are used instead of the approximate scheme
in extreme situations.  We performed additional fits in which the
6 constraints from Table 1 were incorporated. For the $M=N=3$ case (with
10+1 free parameters), the
deviations, denoted by a dagger in Table 2,  are about 4 times larger
than in the unconstrained fit.
Nevertheless, the overall accuracy is still acceptable for
most numerical work, and if more accuracy is desired, one could increase
$M$ or $N$.  Note that for the case $M=N=3$, the improvement in the
pressure error gained by fitting $a$ amounts to a factor of 9.  We
list the coefficients for this case in Table 5.

\section{APPROXIMATION TO BOSON \protect\\INTEGRALS}
\subsection{Limiting Cases}

The approximation to boson integrals follows by analogy to that for
fermions. However, the degenerate case is now denoted by $|\psi|\ll1$ since
$\psi$ must always be negative.  Furthermore, a boson--antiboson gas in
equilibrium has the additional restriction that $\psi t\geq-2$, which would
preclude the existence of the non-degenerate, extremely relativistic case,
although it is nevertheless useful to consider this formal limit below.

We first determine the expansions in the various limiting cases.
\subsubsection{Non-degenerate}
The expression for the pressure in the non-degenerate ($\psi\ll-1$) case
was given in equation (\ref{ndf}) and from this the internal energy density and 
density are easily obtained.

For the non-relativistic case, in which $t\ll1$, the asymptotic series
for the modified Bessel functions yields
\begin{eqnarray}
p&\pj=&\pj \sqrt{\frac{\pi}{2}}t^{\frac{5}{2}}e^{\psi}
\sum\limits_{n=1}^{\infty}\frac{e^{(n-1)\psi}}{n^{\frac{5}{2}}}\left\{
1+\frac{15t}{8n}+\frac{105t^2}{128n^2}-\cdots\right\}\;,\nonumber\\
u&\pj=&\pj \frac{3}{2}\sqrt{\frac{\pi}{2}}t^{\frac{5}{2}}e^{\psi}
\sum\limits_{n=1}^{\infty}\frac{e^{(n-1)\psi}}{n^{\frac{5}{2}}}\left\{
1+\frac{25t}{8n}+\frac{245t^2}{128n^2}-\cdots\right\}\;,\nonumber\\
\rho&\pj=&\pj \sqrt{\frac{\pi}{2}}t^{\frac{3}{2}}e^{\psi}
\sum\limits_{n=1}^{\infty}\frac{e^{(n-1)\psi}}{n^{\frac{3}{2}}}\left\{
1+\frac{15t}{8n}+\frac{105t^2}{128n^2}-\cdots\right\}\;.\qquad{\rm ND\ NR}
\label{ndnr}
\end{eqnarray}

As we have remarked, we consider the extremely relativistic ($t\gg1$) case
even though it is not physical when one is 
dealing with a boson-antiboson gas in thermal equilibrium.
Making the small argument expansion of the modified Bessel functions, one 
obtains
\begin{eqnarray} p&\pj=&\pj
2t^4e^{\psi}\sum\limits_{n=1}^{\infty}\frac{e^{(n-1)\psi}}{n^4}
\left\{1+\frac{n}{t}+\frac{n^2}{4t^2}-\cdots\right\}\;,\nonumber\\
u&\pj=&\pj
6t^4e^{\psi}\sum\limits_{n=1}^{\infty}\frac{e^{(n-1)\psi}}{n^4}
\left\{1+\frac{2n}{3t}+\frac{n^2}{12t^2}-\cdots\right\}\;,\nonumber\\
\rho&\pj=&\pj
2t^3e^{\psi}\sum\limits_{n=1}^{\infty}\frac{e^{(n-1)\psi}}{n^3}
\left\{1+\frac{n}{t}+\frac{n^2}{4t^2}-\cdots\right\}\;.\qquad{\rm ND\
ER} \label{nder} \end{eqnarray}

\subsubsection{Extremely Degenerate}

In the non-relativistic case ($|\psi|\ll1, t\ll1$), we use $y=E/t$ as the 
integration variable in equation (\ref{p}) to find
\begin{equation}
p=\oneth te^{\psi}\int\limits_0^{\infty}dy\:(2yt)^{\frac{3}{2}}\left\{1+
{\textstyle\frac{3}{4}}yt+{\textstyle\frac{3}{32}}(yt)^2-\cdots\right\}
[e^y-e^{\psi}]^{-1}\;.\label{p2}
\end{equation}
The integrals can be written in terms of a function $\Phi$ which is defined 
by Erd\'elyi {\it et al.} (1953) as
\begin{eqnarray}
\Phi(e^{\psi},s,1)&\pj&\pj=[\Gamma(s)]^{-1}\int\limits_0^{\infty}dt\:t^{s-1}
[e^t-e^{\psi}]^{-1}\ \ ;\ \ s>1,\ \psi\leq0\nonumber\\
&\pj=&\pj e^{-\psi}\left\{\Gamma(1-s)(-\psi)^{s-1}
+\sum\limits_{r=0}^{\infty}\zeta(s-r)\frac{\psi^r}{r!}\right\}\;,\label{phi}
\end{eqnarray}
where $\zeta(z)$ is the Riemann zeta function. For the series expansion to 
be valid, $|\psi|<2\pi$ and $s\neq1,2,3\ldots$. This function is related to 
the polylogarithm function employed by Haber \& Weldon (1982a, b);
$\Phi(e^{\psi},s,1)=e^{-\psi}{\rm Li}_s(e^{\psi})$. 
Using equations (\ref{p2}, \ref{phi}) we find
\begin{eqnarray}
p&\pj=&\pj\sqrt{\frac{\pi}{2}}t^{\frac{5}{2}}\left\{
\zeta\left({\textstyle\frac{5}{2}}\right)
+\zeta\left({\textstyle\frac{3}{2}}\right)\psi
+\thalf\zeta\left({\textstyle\frac{1}{2}}\right)\psi^2
+\fourth\sqrt{\pi}(-\psi)^{\frac{3}{2}}
+{\textstyle\frac{15}{8}}\zeta\left({\textstyle\frac{7}{2}}\right)t
\right.\nonumber\\&&\left.
+{\textstyle\frac{15}{8}}\zeta\left({\textstyle\frac{5}{2}}\right)\psi t
+{\textstyle\frac{105}{128}}\zeta\left({\textstyle\frac{9}{2}}\right)t^2
+\cdots\right\}\;,\nonumber\\
u&\pj=&\pj\frac{3}{2}\sqrt{\frac{\pi}{2}}t^{\frac{5}{2}}\left\{
\zeta\left({\textstyle\frac{5}{2}}\right)
+\zeta\left({\textstyle\frac{3}{2}}\right)\psi
+\thalf\zeta\left({\textstyle\frac{1}{2}}\right)\psi^2
+\fourth\sqrt{\pi}(-\psi)^{\frac{3}{2}}
+{\textstyle\frac{25}{8}}\zeta\left({\textstyle\frac{7}{2}}\right)t
\right.\nonumber\\&&\left.
+{\textstyle\frac{25}{8}}\zeta\left({\textstyle\frac{5}{2}}\right)\psi t
+{\textstyle\frac{245}{128}}\zeta\left({\textstyle\frac{9}{2}}\right)t^2
+\cdots\right\}\;,\nonumber\\
\rho&\pj=&\pj\sqrt{\frac{\pi}{2}}t^{\frac{3}{2}}\left\{
\zeta\left({\textstyle\frac{3}{2}}\right)
+\zeta\left({\textstyle\frac{1}{2}}\right)\psi
+\thalf\zeta\left(-{\textstyle\frac{1}{2}}\right)\psi^2
-2\sqrt{\pi}(-\psi)^{\frac{1}{2}}
+{\textstyle\frac{15}{8}}\zeta\left({\textstyle\frac{5}{2}}\right)t
\right.\nonumber\\&&\left.
+{\textstyle\frac{15}{8}}\zeta\left({\textstyle\frac{3}{2}}\right)\psi t
+{\textstyle\frac{105}{128}}\zeta\left({\textstyle\frac{7}{2}}\right)t^2
+{\textstyle\frac{5}{2}}\sqrt{\pi}t(-\psi)^{\frac{3}{2}}
+\cdots\right\}\,.\quad{\rm ED\ NR} \label{ednr}
\end{eqnarray}

The opposite case of extremely relativistic, degenerate bosons
($|\psi|\ll1, t\gg1$) requires a series expansion of the pressure integral,
followed by a Mellin transformation and the evaluation of the residues
at the poles. Using the results of Haber \& Weldon (1982a, b), we find
\begin{eqnarray}
p\pj&=&\pj 2t^4\Biggl\{\zeta(4)+\zeta(3)\alpha
+\zeta(2)\left(\frac{\alpha^2}{2}-\frac{1}{4t^2}\right)
+\frac{11\alpha^3}{36}-\frac{7\alpha}{24t^2}\nonumber\\
&&+\left(\frac{\alpha^3}{6}-\frac{\alpha}{4t^2}\right)\ln(2t)
+\frac{1}{6}\left[\frac{1}{t^2}-\alpha^2\right]^{\frac{3}{2}}
\left[\frac{\pi}{2}+{\rm arcsin}(\alpha t)\right]
+\cdots\Biggr\}\;,\nonumber\\
u\pj&=&\pj 6t^4\Biggl\{\zeta(4)+\zeta(3)\left(\alpha-\frac{1}{3t}\right)
+\zeta(2)\left[\frac{\alpha^2}{2}-\frac{\alpha}{3t}-\frac{1}{12t^2}\right] 
\nonumber\\
&&-\frac{4\alpha^3}{9}+\left(\frac{3\alpha^2}{2}-\frac{1}{4t^2}\right)
\left(\frac{\alpha}{2}-\frac{1}{6t}\right)
+\frac{\psi}{6}\left(\alpha^2-\frac{1}{2t^2}\right)\ln(2t)\nonumber\\
&&-\frac{\psi\alpha}{6}\left[\frac{1}{t^2}-\alpha^2\right]^{\frac{1}{2}}\left[
\frac{\pi}{2}+{\rm arcsin}(\alpha t)\right]+\cdots\Biggr\}\;,\nonumber\\
\rho\pj&=&\pj 2t^3\Biggl\{\zeta(3)+\zeta(2)\alpha
+\frac{3\alpha^2}{4}-\frac{1}{8t^2}
+\left(\frac{\alpha^2}{2}-\frac{1}{4t^2}\right)\ln(2t)\nonumber\\
&&-\frac{\alpha}{2}\left[\frac{1}{t^2}-\alpha^2
\right]^{\frac{1}{2}}\left[\frac{\pi}{2}+{\rm arcsin}(\alpha t)\right]
+\frac{\zeta(0)\alpha}{6}\left(\alpha^2-\frac{3}{2t^2}\right)+\cdots\Biggr\}
\,,\nonumber\\
&& \hspace{9cm} {\rm ED\ ER}\label{eder}
\end{eqnarray}
where $\alpha=\psi+t^{-1}$. Note that $-1\leq\alpha t\leq1$, if 
$-2\leq\psi t\leq 0$.

\subsection{Numerical Approximation}

We now follow the strategy employed for fermions by EFF.
The preceding formulae suggest that for the
non-relativistic (relativistic) case, we need a power series in $t$
($t^{-1}$). Thus, a suitable choice is simply $g=t$.
In the non-degenerate limit, powers of $e^{\psi}$ are
required. For the degenerate case, powers of $\sqrt{-\psi}$ are
indicated by equation (\ref{ednr}); the situation  is less clear for equation 
(\ref{eder}), which has a complicated structure, but it would seem
reasonable to use a similar approach. We are therefore led to  consider
a variable $h$ such that
\begin{eqnarray} 
h\propto e^{-\psi}\ &;&\ \frac{dh}{d\psi}\propto-h,\qquad|\psi|\gg1,
\ h\gg1\quad{\rm ND}\nonumber\\
h\propto \sqrt{-\psi}\ &;&\ \frac{dh}{d\psi}\propto-\frac{1}{h},
\qquad|\psi|\ll1,\ h\ll1\quad{\rm ED}
\end{eqnarray} 
This does not uniquely specify the derivative and, after some
experimentation, we chose
\begin{equation}
\frac{dh}{d\psi}=-\frac{\left(\sqrt{a}+h\right)^2}{h}\;,
\end{equation}
which has the desired limiting behavior. The introduction of the
parameter $a$ allows greatly improved fits, as we have also found in the
fermion case (see \S 3).  This property is related to
the limiting behavior of the number density and entropy in the degenerate
limit.  Integration then yields
\begin{equation}
\psi=\frac{h}{\sqrt{a}+h}-\ln\left(\frac{\sqrt{a}+h}{\sqrt{a}}\right)\;.
\label{psih}
\end{equation}
For small values of $h$ and $\psi$, $h\simeq\sqrt{-2a\psi}$, while for 
large values $h\simeq \sqrt{a}e^{1-\psi}$; note that $h\geq0$ for all
$\psi\leq0$.  

We then write the polynomial expansion for the pressure
\begin{equation}
p=\frac{t^{\frac{5}{2}}(1+t)^{\frac{3}{2}}}{(1+h)^{M+1}(1+t)^N}
\sum\limits_{m=0}^M\sum\limits_{n=0}^Np_{mn}h^mt^n\;,\label{ppoly}
\end{equation}
where the coefficients of the polynomial are denoted by $p_{mn}$.
(Note that in the case $h=0$, the interpretation
$h^m=\delta_{0m}$ is to be made.)
For the ED case, where $h$ and $|\psi|$ are small, $p$ should not contain a 
term proportional to $h$, which would yield a leading divergent  
$1/\sqrt{-\psi}$ contribution to the 
number density. Therefore, we must impose the condition $p_{1n}=(M+1)p_{0n}$ so
that the contribution to equation (\ref{ppoly}) which is linear in $h$ 
vanishes. Equation (\ref{ppoly}) then
has the correct behavior in the four limiting cases and should be 
suitable for smoothly interpolating for all values of $\psi$ and $t$.
Since it is desirable that the internal energy density and number density 
be thermodynamically consistent with the pressure, we differentiate 
equation (\ref{ppoly}) to obtain
\begin{eqnarray}
u&\pj=&\pj\frac{t^{\frac{5}{2}}(1+t)^{\frac{3}{2}}}{(1+h)^{M+1}(1+t)^N}
\sum\limits_{m=0}^M\sum\limits_{n=0}^Np_{mn}h^mt^n
\left[\threehalf+n+(\threehalf-N)\frac{t}{1+t}\right]\;,\nonumber\\
\rho&\pj=&\pj\frac{(h+\sqrt{a})^2[t(1+t)]^{\frac{3}{2}}}{(1+h)^{M+2}(1+t)^N}
\sum\limits_{m=0}^M\sum\limits_{n=0}^Np_{mn}h^{m-2}t^n
[-m+h(M+1-m)].\label{unpoly}
\end{eqnarray}
The entropy density can then be obtained from $s=(p+u)/t-\rho\psi$.  We note
that the condition $p_{1n}=(M+1)p_{0n}$ leads to explicit cancellation of
the leading order contributions to the entropy in the degenerate limit.

Note that equation (\ref{ppoly}) for bosons is actually the same as equation 
(\ref{fppoly}) for fermions as may be seen by setting $f=1/h$ and using the
boson parameterization $g=t$.  It is simply more convenient to use
these forms so that infinite values of $h$ for bosons or $f$ for fermions
are only encountered for infinitely large $|\mu|$.  In the boson case, it
is necessary to be able to specify the case $\psi=0$, which is just $h=0$.

The leading terms for the limiting expressions for $p$ and $\rho$, given
in  \S 4.1, yield values for selected coefficients $p_{mn}$  and these
are  collected in Table 6. For reference, we note that
$\zeta(3/2)=2.61238, \zeta(5/2)=1.34149, \zeta(3)=1.20206$, and
$\zeta(4)=\pi^4/90$. The fitted values for these coefficients  should 
approach the values in Table 6 with increasing dimensions $M$ and $N$,
as indeed we find. For the case $M=2$ we have two different
requirements on the coefficients, derived from the limiting cases for
$p$ and $\rho$. For $p_{20}$ these expressions  agree for $a=0.9486$ or
$a=0.9999$, while for $p_{2N}$ they agree if $a=0.3854$ or $a=0.9005$. 
Using $a=0.91$, for example, the expressions agree to $\leq0.25$\% in both 
cases. On the other hand,
for $a=1$ the expressions for $p_{20}$ agree to 0.0003\%, but the ones
for $p_{N0}$ agree to only 2.7\%.  This large error explains why it
is advantageous to employ $a$ as an additional fitting parameter.

\subsection{Results}

The polynomial coefficients $p_{mn}$ in equations (\ref{ppoly}) 
and (\ref{unpoly}) were obtained by a least squares fit to ``exact" 
results obtained by numerical Gaussian integration (the accuracy was
at least six decimal places). The results were generated for a $(t,\psi)$ 
grid consisting of $t$ values of 
.05, .1, .5, 1, 5, 10 and 100 and, for each $t$, seven equally spaced 
values of $\psi$ in the range $-2\leq\psi t\leq0$. The overall quality of 
the results obtained for various dimensionalities $M$ and $N$ is
indicated in Table 7 in terms of the maximum modulus and 
the root mean square deviations, as for fermions.
If the coefficients $p_{mn}$ are obtained by fitting just the pressure data,
then a comparable fit is obtained for the internal  energy. However the fit
for the number density can be significantly improved by separately 
optimising the $p_{mn}$ coefficients. Therefore the numbers 
in Table 7 were obtained by simultaneously fitting the pressure and number
density data, thus sacrificing some accuracy in $p$ to achieve improvement
in $\rho$ (both the sacrifice and improvement are typically a factor of
$\sim4$ in the RMS deviations). In addition to minimizing with respect to 
the $(N+1)M$
coefficients $p_{mn}$, the results were minimized with respect to the 
parameter $a$ of equation (\ref{psih}). Although $a$ turned out to be fairly 
close to 1, there is a substantial improvement gained by allowing it to
vary. For the cases listed $a$ was in the range $0.7-1.1$.

The simplest approximation that is sensible, $M=2\ N=1$, produces results 
accurate to better than 5\% with only $4+1$ parameters, while the $M=N=2$
case with $6+1$ parameters achieves accuracies of $\lord\thalf$\%. These
cases are sufficiently simple that they could be employed in hand 
calculations to yield reasonable estimates. We therefore tabulate the 
coefficients in Tables 8 and 9, respectively. Increasing the 
number of parameters obviously improves the accuracy and we would judge 
that the $M=3\ N=4$ case is sufficient for most purposes. Here the
MM deviations are $<0.02$\% and the RMS deviations are $\sim0.003$\%.
The $15+1$ parameters are given in Table 10. 

As in the case of
fermions, it might
be preferable, or even necessary, to have an approximation which merges
exactly into the asymptotic limiting cases.  Thus, it is natural to
employ  the values of Table 6 as constraints and fit the remaining
parameters. For $M=3$ and $N=4$ this leaves 9+1 free parameters.  The
deviations, denoted by a dagger in Table 7, are, of course, larger than
in the unconstrained case, but only by a factor of $\lord2$ (the 
degradation was slightly worse for fermions). The
coefficients for this case, tabulated in Table 11, are similar to those
of Table 10 since the errors do not differ greatly.

\section{CONCLUDING REMARKS}

We have generated polynomial approximations to the
thermodynamic integrals which embody and merge exactly into the analytic
limiting cases and accurately interpolate over the entire range of
temperature and chemical potential. The boson case had not been treated 
previously and we have improved upon the parameterization of EFF
and the modification of Pols et al. (1995)
for fermions. With 10 (11) free parameters, and 20 (16) terms, we
achieve accuracies of better than 0.008\% (0.03\%) for the boson (fermion)
pressure.   Some simpler, but less accurate, approximations were
also discussed.

In astrophysics, it is often the case that particles and antiparticles are in
thermal equilibrium.  Then, the net number density is ${\cal N}={\cal
N}_+-{\cal N}_-$ where ${\cal N}_+({\cal N}_-)$ is the number density of
particles (antiparticles), and the total pressure is
$P=P_++P_-$.  In thermal equilibrium, we must have $\mu_+=-\mu_-$.  The problem
usually posed is such that one must determine the thermodynamic quantities
given ${\cal N}$ and $T$.  An obvious approach would be to employ the EFF
scheme and to establish the limits of the pressure in analogy to
equation (\ref{flimits}) in the fermion case.  Unfortunately, the 
non-degenerate,
non-relativistic case has no single limit in terms of power laws because of
pair formation.  This is true in both the fermion and boson cases.  Thus, the
EFF scheme cannot be directly applied to pairs in equilibrium.

It is necessary, therefore, to treat pairs in terms of the particles and
antiparticles separately.  Thus, we can use the results of \S 3 for
the number densities ${\cal N}_+(f_+,T)$ and ${\cal N}_-(f_-,T)$ of
fermions and antifermions,
respectively.  Equation (\ref{psif}) gives $\psi_\pm$ in terms of $f_\pm$. The 
pair of simultaneous equations
\begin{equation}
{\cal N}={\cal N}_+(f_+,T)-{\cal N}_-(f_-,T)\quad;\quad
\psi_+(f_+)=-\psi_-(f_-)-2mc^2/T\label{pairs}\;,
\end{equation}
is then solved for $f_+$ and $f_-$, as opposed to the inversion of a single
equation ${\cal N}={\cal N}(f,T)$ for $f$.  For bosons, one simply replaces
$f_\pm$ by $h_\pm$ in equation (\ref{pairs}).

This work was supported in part by the U.S. Department of Energy under
contract numbers DE-FG02-87ER40328 and DE-FG02-87ER40317.

\newpage
\begin{quote} {\bf Table 1.} Fermion coefficients, $p_{mn}$,
derived from the leading terms in the limiting expressions for $p$ and $s$.
\end{quote}
\begin{center}
\begin{tabular}{ccc} \hline \hline $p_{mn}$&$n=0$&$n=N$\\ \hline
$m=0$&$\frac{e^2}{a}\sqrt{\frac{\pi}{32}}$&$\frac{e^2}{2a}$\\ \hline
$m=M-1$&$\frac{1}{15}a^{-\frac{5}{4}}[5\pi^2a+8(4M-5a-1)]$&
$\frac{2}{3a^2}[a(\pi^2-4)+2(M-1)]$\\ \hline
$m=M$&$\frac{32}{15}a^{-\frac{5}{4}}$&$\frac{4}{3a^2}$\\ \hline\hline
\end{tabular}
\end{center}
\vfill
\begin{quote} {\bf Table 2.} Maximum modulus (MM) and root mean
square (RMS) deviations for the fermion thermodynamic quantities for
various dimensionalities $M$ and $N$. \end{quote}
\begin{center}
\begin{tabular}{ccllllllll} \hline \hline
&&\multicolumn{6}{c}{$a$ fitted}&\multicolumn{2}{c}{$a=1$}\\ \cline{3-8}
&&\multicolumn{2}{c}{$p$}&\multicolumn{2}{c}{$u$}&
\multicolumn{2}{c}{$\rho$}&\multicolumn{2}{c}{$p$}\\ 
$M$&$N$&MM&RMS&MM&RMS&MM&RMS&MM&RMS\\ \hline
2&1&2.5($-$2)&9.5($-$3)&3.1($-$2)&1.0($-$2)&2.5($-$2)&9.1($-$3)&2.5($-$2)
&1.0($-$2)\\
2&2&8.1($-$4)&3.4($-$4)&8.0($-$4)&3.3($-$4)&9.1($-$4)&3.6($-$4)&2.1($-$3)
&7.6($-$4)\\
3&2&5.9($-$4)&2.1($-$4)&5.2($-$4)&1.9($-$4)&5.9($-$4)&2.0($-$4)&7.4($-$4)
&3.1($-$4)\\
2&3&7.0($-$4)&2.7($-$4)&7.0($-$4)&2.6($-$4)&9.0($-$4)&3.0($-$4)&1.9($-$3)
&7.1($-$4)\\
3&3&7.1($-$5)&2.6($-$5)&1.3($-$4)&4.7($-$5)&1.3($-$4)&4.0($-$5)&5.6($-$4)
&1.8($-$4)\\
4&3&7.2($-$5)&2.3($-$5)&1.3($-$4)&4.6($-$5)&7.2($-$5)&2.4($-$5)&1.9($-$4)
&7.4($-$5)\\ 
3&4&4.0($-$5)&1.2($-$5)&3.8($-$5)&1.2($-$5)&1.2($-$4)&3.3($-$5)&5.4($-$4)
&1.8($-$4)\\\hline
$\dagger$3&3&3.0($-$4)&8.1($-$5)&4.5($-$4)&1.0($-$4)&3.0($-$4)&8.7($-$5)
&2.6($-$3)&8.6($-$4)\\
\hline\hline
\end{tabular}
\end{center}
\noindent The notation $5.9(-4)$, for example, indicates $5.9\times10^{-4}$.

\noindent$\dagger$ Employing the constraints of Table 1.

\newpage
\begin{quote} {\bf Table 3.} Fermion coefficients $p_{mn}$
for $M=N=2$;  $a=0.442$. \end{quote}
\begin{center}
\begin{tabular}{cccc} \hline \hline $p_{mn}$&$n=0$&$n=1$&$n=2$ \\\hline
$m=0$& 5.23810&12.4991&8.36058 \\ \hline
$m=1$&11.1751&25.3687&15.6453 \\ \hline
$m=2$& 5.91800&12.4945&6.82530 \\
\hline\hline \end{tabular}
\end{center}
\vfill
\begin{quote} {\bf Table 4.} Fermion coefficients $p_{mn}$
for $M=N=3$;  $a=0.420$. \end{quote}
\begin{center}
\begin{tabular}{ccccc} \hline \hline
$p_{mn}$&$n=0$&$n=1$&$n=2$&$n=3$ \\ \hline
$m=0$&5.51219&18.6215&22.0078&8.71963 \\ \hline
$m=1$&17.4343&57.7188&66.1098&25.6323 \\ \hline
$m=2$&18.1239&58.7781&65.1429&24.4246 \\ \hline
$m=3$&6.30952&19.8967&21.1375&7.55866 \\ \hline\hline
\end{tabular}
\end{center}
\vfill
\begin{quote} {\bf Table 5.} Fermion coefficients $p_{mn}$
for $M=N=3$ employing the constraints of Table 1; $a=0.433$.
\end{quote}
\begin{center}
\begin{tabular}{ccccc} \hline \hline
$p_{mn}$&$n=0$&$n=1$&$n=2$&$n=3$ \\ \hline
$m=0$&5.34689&18.0517&21.3422&8.53240 \\ \hline
$m=1$&16.8441&55.7051&63.6901&24.6213 \\ \hline
$m=2$&17.4708&56.3902&62.1319&23.2602 \\ \hline
$m=3$&6.07364&18.9992&20.0285&7.11153 \\ \hline\hline
\end{tabular}
\end{center}
\newpage
\begin{quote} {\bf Table 6.} Boson coefficients, $p_{mn}$,
derived from the
leading terms in the limiting expressions for $P$ and $\rho$.\end{quote}
\begin{center}
\begin{tabular}{ccc} \hline \hline
$p_{mn}$&$n=0$&$n=N$\\ \hline
$m=0$&$\sqrt{\frac{\pi}{2}}\zeta\left({\textstyle\frac{5}{2}}\right)$&
$2\zeta(4)$\\ \hline
$m=1$&$\sqrt{\frac{\pi}{2}}(M+1)\zeta\left({\textstyle\frac{5}{2}}\right)$&
$2(M+1)\zeta(4)$\\ \hline
$m=2$&$\sqrt{\frac{\pi}{8}}\left[-a^{-1}\zeta\left({\textstyle\frac{3}{2}}
\right)+M(M+1)\zeta\left({\textstyle\frac{5}{2}}\right)\right]$&
$-a^{-1}\zeta(3)+M(M+1)\zeta(4)$\\ \hline
$m=M$&$e\sqrt{\frac{a\pi}{2}}$&$2e\sqrt{a}$\\ \hline\hline \end{tabular}
\end{center}
\vfill
\begin{quote} {\bf Table 7.} Maximum modulus (MM) and root
mean square (RMS) deviations for the boson thermodynamic quantities for
various dimensionalities $M$ and $N$. \end{quote}
\begin{center}
\begin{tabular}{ccllllll} \hline \hline
&&\multicolumn{2}{c}{$p$}&\multicolumn{2}{c}{$u$}&
\multicolumn{2}{c}{$\rho$}\\  $M$&$N$&MM&RMS&MM&RMS&MM&RMS\\ \hline
2&1&2.1($-$2)&1.3($-$2)&2.3($-$2)&1.3($-$2)&4.4($-$2)&1.5($-$2)\\
2&2&2.7($-$3)&1.1($-$3)&2.9($-$3)&1.2($-$3)&5.7($-$3)&1.7($-$3)\\
3&2&3.9($-$4)&1.2($-$4)&4.6($-$4)&1.9($-$4)&3.0($-$3)&7.5($-$4)\\
2&3&2.6($-$3)&7.9($-$4)&2.7($-$3)&1.0($-$3)&4.9($-$3)&1.4($-$3)\\
3&3&1.1($-$4)&5.8($-$5)&2.0($-$4)&8.4($-$5)&5.7($-$4)&1.6($-$4)\\
4&3&1.1($-$4)&3.7($-$5)&2.4($-$4)&5.8($-$5)&5.9($-$4)&1.5($-$4)\\ 
3&4&4.9($-$5)&1.7($-$5)&1.3($-$4)&3.5($-$5)&1.7($-$4)&4.7($-$5)\\\hline 
$\dagger$3&4&7.8($-$5)&2.4($-$5)&1.7($-$4)&4.3($-$5)&2.4($-$4)&9.0($-$5)\\
\hline\hline 
\end{tabular}
\end{center}
\noindent$\dagger$ Employing the constraints of Table 6.

\newpage
\begin{quote} {\bf Table 8.} Boson coefficients $p_{mn}$ for
$M=2$ and $N=1$;  $a=0.978$. \end{quote}
\begin{center}
\begin{tabular}{ccc} \hline \hline $p_{mn}$&$n=0$&$n=1$\\ \hline
$m=0$&1.63146&2.11571\\ \hline
$m=1$&\multicolumn{2}{c}{$p_{1n}=3p_{0n}$}\\ \hline
$m=2$&3.31275&5.15372\\ \hline\hline
\end{tabular}
\end{center}
\vfill
\begin{quote} {\bf Table 9.} Boson coefficients $p_{mn}$ for
$M=N=2$;  $a=0.914$.
\end{quote}
\begin{center}
\begin{tabular}{cccc} \hline \hline $p_{mn}$&$n=0$&$n=1$&$n=2$\\
\hline $m=0$&1.68131&3.47558&2.16582\\ \hline
$m=1$&\multicolumn{3}{c}{$p_{1n}=3p_{0n}$}\\ \hline
$m=2$&3.25053&7.82859&5.19126\\ \hline\hline
\end{tabular}
\end{center}
\vfill
\begin{quote} {\bf Table 10.} Boson coefficients $p_{mn}$ for
$M=3,\ N=4$; $a=1.029$.
\end{quote}
\begin{center}
\begin{tabular}{cccccc} \hline \hline
$p_{mn}$&$n=0$&$n=1$&$n=2$&$n=3$&$n=4$\\ \hline
$m=0$&1.68134&6.85070&10.8537&7.81843&2.16461\\ \hline
$m=1$&\multicolumn{5}{c}{$p_{1n}=4p_{0n}$}\\ \hline
$m=2$&8.49651&35.6058&57.7134&42.3593&11.8199\\ \hline
$m=3$&3.45614&15.1152&25.5254&19.2745&5.51757 \\ \hline\hline
\end{tabular}
\end{center}
\newpage
\begin{quote} {\bf Table 11.} Boson coefficients $p_{mn}$ for
$M=3,\ N=4$ including the constraints of Table 6; $a=1.040$. \end{quote}
\begin{center}
\begin{tabular}{cccccc} \hline \hline
$p_{mn}$&$n=0$&$n=1$&$n=2$&$n=3$&$n=4$\\ \hline
$m=0$&1.68130&6.85060&10.8539&7.81762&2.16465\\ \hline
$m=1$&\multicolumn{5}{c}{$p_{1n}=4p_{0n}$}\\ \hline
$m=2$&8.51373&35.6576&57.7975&42.4049&11.8321\\ \hline
$m=3$&3.47433&15.1995&25.6536&19.3811&5.54423 \\ \hline\hline
\end{tabular}
\end{center}
\newpage
\section*{REFERENCES} 

\noindent Abramowitz, M., and  Stegun, I.A. 1965,
eds. Handbook of Mathematical Functions (Dover, NY).\hfill\\

\noindent Chandrasekhar, S. 1958, An Introduction to Stellar Structure
(Dover, NY).\hfill\\

\noindent Eggleton, P.P., Faulkner J., \& Flannery, B.P. 1973, A \& A
23, 325.\hfill\\

\noindent Erd\'elyi, A., Magnus, W., Oberhettinger, F., \& Tricomi, F.G.
1953,  Higher Transcendental Functions (McGraw-Hill, NY) Vol. I, p.27
ff.\hfill\\

\noindent Haber, H.E., \& Weldon, H.A. 1982a,  Phys. Rev.D, 25,
502.\hfill\\

\noindent \rule[1mm]{.7cm}{.2pt} 1982b, J. Math. Phys. 23, 1852.\hfill\\

\noindent Pols, O.R., Tout, C.A., Eggleton, P.P. \& Han, Z. 1995, Mon.
Not. R. Astron. Soc. 274, 964.
\hfill\\

\end{document}